# Evolution of Scientific Collaboration Network Driven by Homophily and Heterophily

Peng Liu, Shuangling Luo, Haoxiang Xia

(School of Management Science and Engineering, Dalian University of Technology, Linggong Road 2, Dalian, Liaoning 116024, China)

**Abstract**  Many scientific collaboration networks exhibit clear community and small world structures. However, the studies on the underlying mechanisms for the formation and evolution of community and small world structures are still insufficient. The mechanisms of homophily and heterophily based on scholars' traits are two important factors for the formation of community and inter-communal links, which may deserve further exploration. In this paper, a multi-agent model, which is based on combinatorial effects of homophily and heterophily, is developed to investigate the evolution of scientific collaboration networks. The simulation results indicate that agents with similar traits aggregate to form community by homophily, while heterophily plays a major role in the formation of inter-communal links. The pattern of network evolution revealed in simulations is essentially consistent with what is observed in empirical analyses, as in both cases the giant component evolves from a small cluster to a structure of "chained-communities", and then to a small world network with community structure. This work may provides an alternative view on the underlying mechanisms for the formation of community and small world structures, complementary to the mainstream view that the small-world is generated from the combination of the structural embeddedness and structural holes mechanisms.

## 1 Introduction

Many recent studies indicate collaboration has become the dominant mode of scientific research in many disciplines (Cronin 2005; Moody 2004; Wuchty et al. 2007). Correspondingly, the study of scientific collaboration has gradually attracted the attention of scholars in computer science, sociology, management and other fields (Cronin 2005; Wuchty et al. 2007; Gazni et al. 2012; Wagner et al. 2002). It has long been realized that the co-authorship of articles in journals provides a window on patterns of collaboration within the academic community. Co-authorship of a paper can be thought of as documenting a collaboration between two or more authors (Newman 2004). With the development of complex network studies in recent years, the scientific collaboration networks also have been extensively investigated in the field of non-linear science (Barabási et al. 1999; Newman 2001; Guimerà 2005). Especially, the WS small-world model (Watts and Strogatz 1998) and BA scale-free model (Barabási and Albert 1999) inspire intensive explorations of such two characteristics in the co-authorship network (Barabási et al. 1999; Newman 2001; Liu et al. 2005; Tomassini and Luthi 2007; Perc 2010; Uzzi et al. 2007). Barabási et al. (1999) and Newman (2001) uncovered the "small-world" and "scale-free" features of co-authorship networks through analyzing the data in the disciplines of mathematics, neuro-science, physics, biomedical sciences and computer science. By summarizing the works on small-world networks in management science, Uzzi et al. (2007) found many co-authorship networks have small-world structure with high clustering coefficient and low average-shortest-path-length. Besides, community or modular structure is another hotspot of the structure in co-authorship networks. By examining collaboration of scholars in Santa Fe Institute, Newman and Girvan (2004) found the distribution of coauthoring relationships is nonuniform and the scholars with dense relationships form communities in the co-authorship network. Evans et al.'s (2011) work also reveals significant community structure of co-authorship networks in the field of business and management in RAE (Research Assessment Exercise).

These above works arouse the scholars' further investigation of the mechanisms that underlie structural evolution of scientific collaboration network by dynamic models. The study on evolutionary dynamics of scientific collaboration network is an important part of exploring social network evolution. Hence, the studies of evolutionary mechanisms in social network may provide some good references for exploring the evolutionary mechanisms of scientific collaboration networks. In the existing models, scale-free feature is mostly attributed to explicit or implicit preferential attachment (Barabási et al. 1999); for the formations of community and small-world networks, people usually proffer an explanation by the combination of structural embeddedness and structural holes. Structural embeddedness (Granovetter 1985) indicates friends' friend is also friend, which establishes the strong tie between individuals, and further causes the formation of communities. While structural embeddedness provides insight to the inner-community links, it does not account for the formation of inter-community links. Explanation for such links rely on another social mechanism of structural holes (Burt 1992), which is often formed by weak ties. Inter-community links create short-cut for individuals in different communities that leads to the formation of small world networks. The combination of such two mechanisms provides a powerful explanation for the formations of community structure and small world networks (Baum et al. 2003; Uzzi and Spiro 2005). For the scientific collaboration networks, there are also evidences that structural embeddedness and structural holes have significant impacts of network evolution. Through analyzing the relationship between individuals in DBLP (Digital Bibliography & Library Project), Backstrom et al. (2006) found fast growth of community size is positive correlated with structural embeddedness among the neighbors. Under the combination of structural embeddedness and randomly creating ternary relationships, Lee et al.'s (2010) model also exhibited the generated network has similar structure to real scientific collaboration networks – networks evolve from the segregation state to the aggregation state (i.e. community formation), then to the structure of large loops (i.e. small world networks).

The prior research endeavors reveal that structural embeddedness and structural holes may be critical mechanisms for the evolution of collaboration networks and the formation of small-world structure. However, in addition to such two mechanisms based on social capital, the establishment of collaborative relationship may also be relevant to knowledge and specialty background of scholars. For example, Newman and Girvan (2004), and Evans et al. (2011) respectively found scholars have similar research topics and specialties in the same community. Such mechanism that similar individuals are more likely to form relationships is called homophily in social science (McPherson et al. 2001). Hence, specialized knowledge similarity is an essential prerequisite of successful collaboration, and it is reasonable to assume that homophily has a significant impact on the formation of scientific collaboration. Meanwhile, the collaborations between researchers with diverse knowledge backgrounds are also a commonplace. For example, quantitative work is more likely to facilitate heterogeneous collaboration (Moody 2004) which is helpful to solve complex problems (Page 2008; Börner et al. 2010). Guimerà et al.'s work (2005) also indicates the first collaboration of scholars from different institutions is easier to produce high quality works. These works show that scholars also have a tendency to collaborate with heterogeneous ones, namely heterophily (Rogers and Bhowmik 1970). Therefore, besides structural embeddedness and structural holes, homophily and heterophily also have a remarkable effect on the evolution of scientific collaboration networks. But, although modeling studies on homophily and heterophily have made some beneficial results (Kimura and Hayakawa 2008; Luo et al. 2015), the evolutionary dynamics of scientific collaboration networks driven by the combinatorial effects of such two mechanisms may deserve further exploration. Especially, homophily is an aggregative mechanism which is good for community formation; and heterophily breaks local aggregation and facilitates inter-community links. The combination of such two mechanisms in some way may promote scientific collaboration network evolves into a small-world network with community structure. The combinatorial effects of homophily and heterophily thus provide a complementary explanation of small-world formation for structural embeddedness and structural holes.

In order to examine the evolutionary mode and underlying mechanism of scientific collaboration networks, the authors of this paper have investigated the co-authorship network of the interdisciplinary field of "evolution of cooperation" (the "EOC network" for short) in a previous work (Liu and Xia 2015). The results indicate the giant component of EOC co-authorship network evolves from a small cluster to a chained-community structure, then to a modular and cohesive small-world network. Meanwhile, the trend of community-research-topic convergence is inherently correlated with network evolution, which indicates homophily may be an important factor to promote the formation of community in the examined network. This paper is a parallel study of such empirical work. In this paper, we develop a multi-agent model to investigate the evolution of scientific collaboration network driven by the combinatorial effects of homophily and heterophily, and then qualitatively compare our results with different evolutionary stages observed in EOC network to further detect whether the presented model reflects the evolutionary mode of real network. The remainder of this paper is organized as follows. In section 2, we briefly describe the evolutionary characteristics of EOC network. In section 3, the multi-agent model is proposed. Subsequently, in section 4, we analyze the simulation results and make a comparison of the EOC co-authorship network and the generated network. At last, the whole paper is concluded in section 5.

## 2 An example of co-authorship network evolution

As mentioned in the introduction section, in a previous work, we have examined the evolution of co-authorship network in the EOC field. We retrieve literatures on EOC from core collection of Web of Science (WoS) and Google Scholar during 1945~2013. We obtain 2583 papers written by 3670 authors, and then construct the cumulative co-authorship network at one year interval. Through examining the cumulative structure of EOC co-authorship network, we find with the size expansion, this network gradually evolves from a segregation state to a core-periphery structure, and such core (i.e. the giant component of EOC co-authorship network) exhibits three different structures. Figure 1 illustrates the giant-component topology of cumulative co-authorship network in typical years. Up to 1999, the giant component was actually a cluster composed of only 20 authors (the left part of Figure 1). And then to 2006, as shown in the middle part of Figure 1, the giant component formed a chained-community structure, which was comprised of three communities interlinked one by one so that the distance between two random nodes could be large.

With further expansion of the giant component till 2013, as shown in the right part of Figure 1, we observe local communities were interlinked by short-cut edges, and hence a small world emerged in the giant component. As a whole Figure 1 shows the process in which the giant component evolves from a cluster to chained-community structure, and then to a small world network. Such evolutionary process is similar to the results in Lee et al.'s (2010) work when they examined the co-authorship network of "complex network research" in theoretical physics.

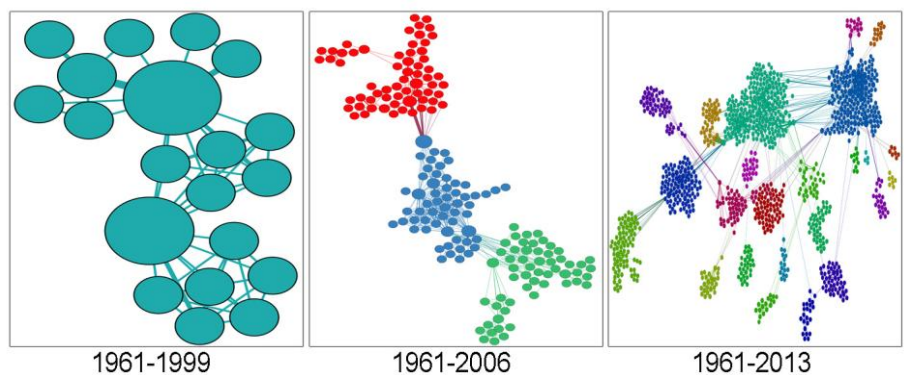

**Figure 1** Topological structures of the giant component of cumulative co-authorship network in typical years (from (Liu and Xia 2015))

By tracking the variation of key topical terms, we examine the relationship between co-authorships and research topical terms. Figure 2 shows the key topical terms of each community in the giant component till 2010, where the circle represents a community, the size of each circle denotes the amount of authors, and text lists the topical terms of each community. In Figure 2, the giant component has a small-world structure in which the communities are interlinked by sparse edges. Through comparative analysis of key topical terms in each community, we find the key topical terms of authors in the same community are similar to each other, and dissimilar to that of authors in other communities. For example, community 387 mainly concerns spatial game, especially the combination of EOC and complex network. The topical terms of community 620 contain "reputation" and "social norm", reveals authors of this community study EOC from the view of social and behavioral science. This indicates the community arises from intensive collaboration of authors whose research topics are similar. In addition, as shown in Figure 3, we find the amount of non-same-community neighbors is roughly proportional to the degree ($k$), which implies high-degree author prefers to take heterogeneous individuals into teamwork. Based on the above examination, we can basically ascertain that both homophily and heterophily play an important role in the evolution of scientific collaboration network. However, such evolutionary process (i.e. from the segregation state to a chained-community structure, then to a small world network) still deserves further exploration. In particular, whether and how the combination of homophily and heterophily prompt scientific collaboration network evolves into a small world network. Hence, our aim in this paper is to model the evolution of scientific collaboration network driven by homophily and heterophily and, where possible, to qualitatively compare our results with different evolutionary stages observed in EOC network.

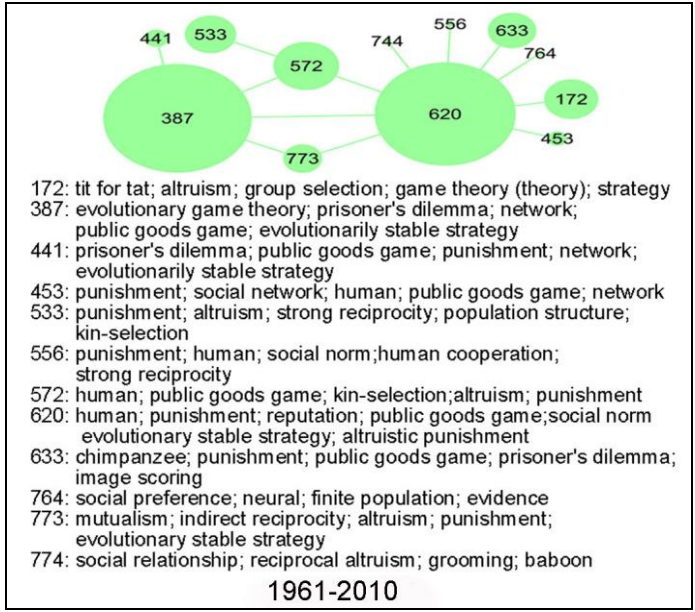

**Figure 2**   The main research topics of each community within the giant component(1961~2010) (from (Liu and Xia 2015))

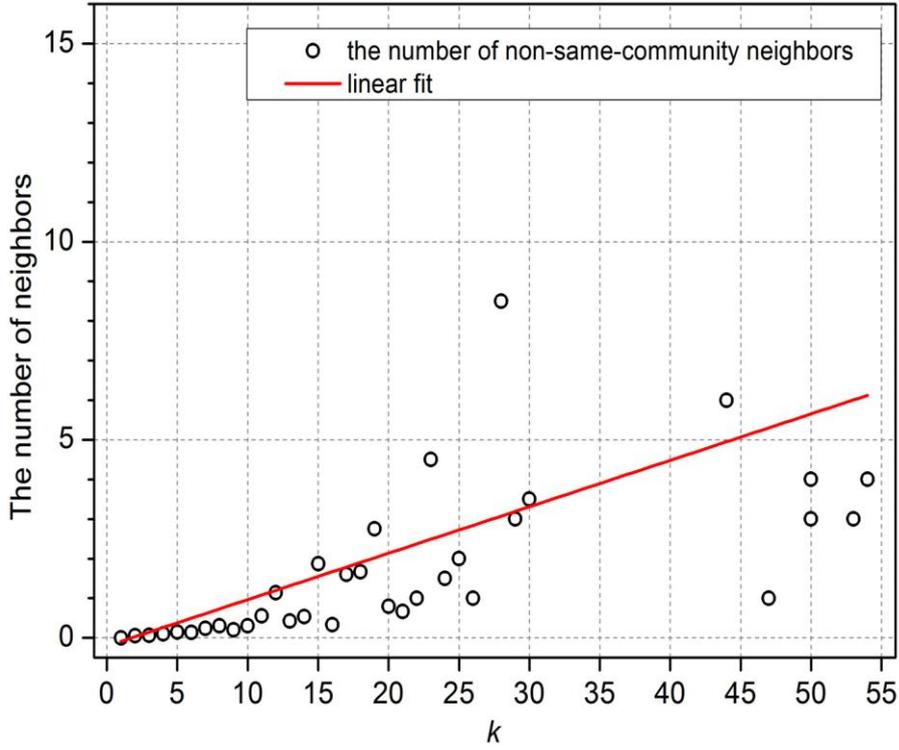

**Figure 3**   The number of other-community-neighbors of specific degree nodes in the giant component (1961~2013) and linear fitting

## 3   The model

Nowadays more and more scientific inquiries are accomplished through teamwork (Moody 2004; Wuchty et al. 2007), i.e. a scientific activity involves two or more scholars. Furthermore, the scientific collaboration team always needs an originator, who plays an important role in team member selection. In the partner-selection process, scholars not only tend to collaborate with similar ones, i.e. homophily; but also choose a heterogeneous scholar as a partner, i.e. heterophily. Existing empirical studies show that such two different tendencies do not occur simultaneously, especially, the mechanism of heterophily requires a certain prerequisite. Scholl (1996) claimed that effective teams require the involvement of individuals with heterogeneous knowledge, and perform well when the proportion of heterogeneous members is in a certain range. By analysis of 14 million literatures from Web of Science, Gazni et al. (2012) found the largest teams have become more diverse than the latter teams. Both the two works indicate heterophily often plays a role in the formation of large scale teams. Furthermore, as described in section 2, we find high-degree authors are much easier to collaborate with the individuals of different specialized knowledge (Figure 3). Therefore, in the modeling process, we introduce a heterogeneous-neighbor-ratio to control team originator whether adds another heterogeneous individual to the team in the next round of collaboration.

Based on the above idea, we borrow Guimerà et al.'s work (2005) to model the evolution of scientific collaboration networks. We firstly consider two types of agents, i.e. incumbents and newcomers, and the newcomers turn into incumbents after taking part in scientific activities. With continued engagement of newcomers, the generated network thus grows and evolves through the formation of new teams. Second, each agent holds a "*trait*", which is an abstraction of scientific characteristics such as specialized knowledge. Third, we assume the originator of each team preferentially selects potential partners by trait homophliy, and then tend to draw a dissimilar agent (i.e. the agent's trait is different from that of the originator) in team when the homogeneity of neighborhood is too strong. By such combinational effects of homophily and heterophily, the agents are interconnected with one another to form teams. In order to avoid excessive concentration of collaboration which may affect the accuracy of simulation results, several teams are generated at each time step in our model. Hence, we introduce a parameters *n-team*, the upper limit of the number of new formed teams in each time step. The assembly of each team is controlled by three parameters: $m$, $p$, and $q$. The first parameter $m$ is the number of team members. In the simulation, the value of $m$ is a positive integer randomly picked from a certain range. The second parameter, $p$, is the probability of a team member being a newcomer. Higher values of $p$ imply more chance for newcomers to enter a field. The third parameter, $q$, is a threshold of the heterogeneous-neighbor-ratio. The overall procedure of the proposed model is comprised of the following steps.

(1) The initial collaboration network is comprised of several isolate nodes (i.e. agents), and these agents are all incumbents.

(2) No more than *n-team* new teams are generated at each time step. In the formation of each new team, the following sub-steps are processed.

In each new team, (2.1) with probability $p$, arbitrarily select a newcomer as the team originator, otherwise (i.e. with probability 1‐$p$), arbitrarily select an incumbent as the team originator.

(2.2) With probability $p$, the team originator randomly selects a newcomer with same trait as a team member and, with complementary probability 1‐$p$, an incumbent is added. When selecting an incumbent as a team member, the team originator respectively counts the number of heterogeneous neighbors and overall neighbors after adding another incumbent with different trait to its neighbors, and then computes the ratio of such two numbers (i.e.

heterogeneous-neighbor-ratio = the number of heterogeneous neighbors/the number of overall neighbors). If this ratio is greater than the threshold $q$, an incumbent whose trait is same to that of the originator is randomly selected; otherwise (i.e. the heterogeneous-neighbor-ratio is not greater than the threshold $q$), arbitrarily add an incumbent with different trait.

It should be noted that the team assembly will be failed when the team originator cannot find enough eligible team members (i.e. the number of team members is less than $m$). Thus, in each time step, the number of new teams will not be always equal to the value of *n-team*.

(3) Repeat step (2) until the count of time steps reaches the pre-specified upper limit.

To study the structure and evolution of the generated network, for different values of the parameters $p$ and $q$, we consider the time-step evolution of several statistical quantities in the network. We use normalized size of the giant component $S(t) = S_g(t)/S_n(t)$ (Guimerà et al. 2005) to measure the connectivity of network, where $S_g(t)$ is the size of the giant component and $S_n(t)$ denotes the size of overall network at time step $t$. The modularity $Q(t)$ (Blondel et al. 2008) is adopted to detect the community structure of the generated network. Furthermore, the small world characteristics are investigated by $C(t) = C_g(t)/C_r(t)$ and $L(t) = L_g(t)/L_r(t)$ (Watts and Strogatz 1998; Uzzi and Spiro 2005), where $C_g(t)$ is the clustering coefficient of the giant component in the generated network, $L_g(t)$ is the average shortest path length of the giant component, $C_r(t)$ and $L_r(t)$ are the counterparts of a same-scale random network.

## 4 Simulations and result analysis

### 4.1 The Settings of numerical experiment

By the proposed model, we conduct numerical experiments to study the structure and evolution of scientific collaboration networks, and the parameter settings are listed in Table 1. In Table 1, in order to avoid excessive concentration of collaborations, the initial collaboration network is comprised of several incumbents (i.e. *ini-incumbent*), and several new teams form at each time step (i.e. *n-team*). Because this paper mainly concerns the evolution of scientific collaboration networks driven by homophily and heterophily, we just consider that the range of trait values is large enough for selecting different heterogeneous partners in team assembly. The setting of team size has consulted the number of authors per paper observed in empirical studies (Newman 2001; Gazni et al. 2012). Subsequently, we analyze the simulation results under different combinations of $p$ and $q$ values.

**Table 1** Setting the simulation conditions

| parameter | value | description |
|---|---|---|
| *ini-incumbent* | 4 | the number of incumbents who comprise the initial collaboration network |
| *n-team* | random integer ranges from 1 to 5 | the number of new teams at each time step |
| *trait* | random integer ranges from 0 to 9 | the abstraction of scientific characteristics, different values mean the scientific characteristics of agents is different |
| *m* | random integer ranges from 1 to 4 | the size of each team |
| $p$ | the value increases from 0 to 1.0 at 0.1 interval | the probability of adding a newcomer to team |
| $q$ | the value increases from 0 to 1.0 at 0.1 interval | the threshold of heterogeneous-neighbor-ratio |
| upper limit of time step | $10^3$ | |

## 4.2 Simulation results and its analysis

### 4.2.1 The structures of the generated network and its giant component

Figure 4 shows the structural characteristics of the generated network and that of its giant component when time step reaches $10^3$. As shown in Figure 4(a), because the value of $S(10^3)$ is less than 1.0, the overall network is in a segregation state. Nevertheless, when $0.1 \leq p \leq 0.5$ and $0.1 \leq q \leq 1.0$ (i.e. the range surrounded by black solid line), the generated network is basically connected due to $S(10^3) \geq 0.8$. In other words, in this range, there is a large giant component which scale is close to that of the entire network. Thus, for the structures of such giant component, especially when $0.1 \leq p \leq 0.5$ and $0.1 \leq q \leq 1.0$, we respectively examine its modularity ($Q(10^3)$), clustering coefficient ($C(10^3)$), and average shortest path length ($L(10^3)$) in Figure 4(b)-(d). Figure 4(b) shows that, in the range surrounded by the black solid line, the value of $Q(10^3)$ is greater than 0.7, which indicates the giant component has a clear community structure. Comparing with the random network of same scale, when $0.2 \leq p \leq 0.6$ and $0.1 \leq q \leq 1.0$, the giant component has a higher cluster coefficient and its shortest path length is close to that of the random network (e.g. in Figure 4(c) and (d), when $0.2 \leq p \leq 0.5$ and $0.1 \leq q \leq 1.0$, $34 < C < 374$, $1.0 < L < 2.0$). In this sense, the giant component is a small world network as characterized by Watts and Strogatz (1998). Furthermore, Uzzi and Spiro (2005) claimed that the level of connectivity becomes higher as the characteristics of small world network strengthen.

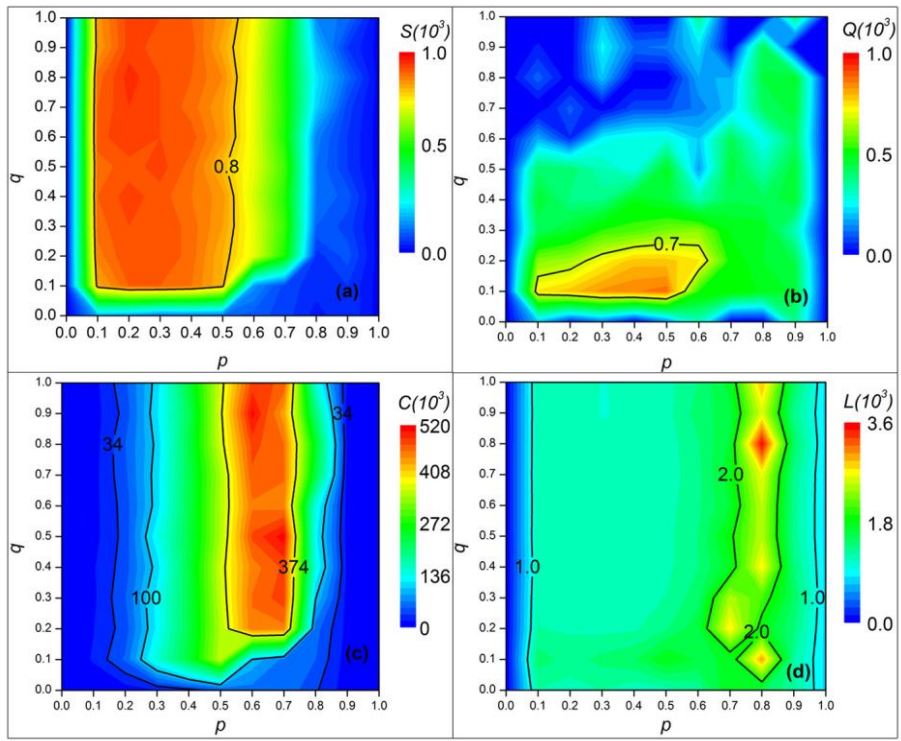

**Figure 4** Structural characteristics of the generated network and that of its giant component (Time step=$10^3$): at different combination of $p$, $q$ values, (a) the size of the giant component (i.e. connectivity of the generated network, denoted as $S(10^3)$), (b) the modularity of the giant component (denoted as $Q(10^3)$), (3) clustering coefficient of the giant component/that of the same-scale random network (denoted as $C(10^3)$), (d) average shortest path length of the giant component/that of the same-scale random network (denoted as $L(10^3)$)

Hence, by further examining the changes of $S(10^3)$, $Q(10^3)$, $C(10^3)$, and $L(10^3)$ values in Figure 4, when $0.1 \leq p \leq 0.5$ and $0.1 \leq q \leq 0.2$, the giant component both has significant small-world characteristics and clear community structures. From the above analysis, we can basically assert that the combinational effects of homophily and heterophily can promote the formations of community structure and small world network in the giant component of scientific collaboration networks.

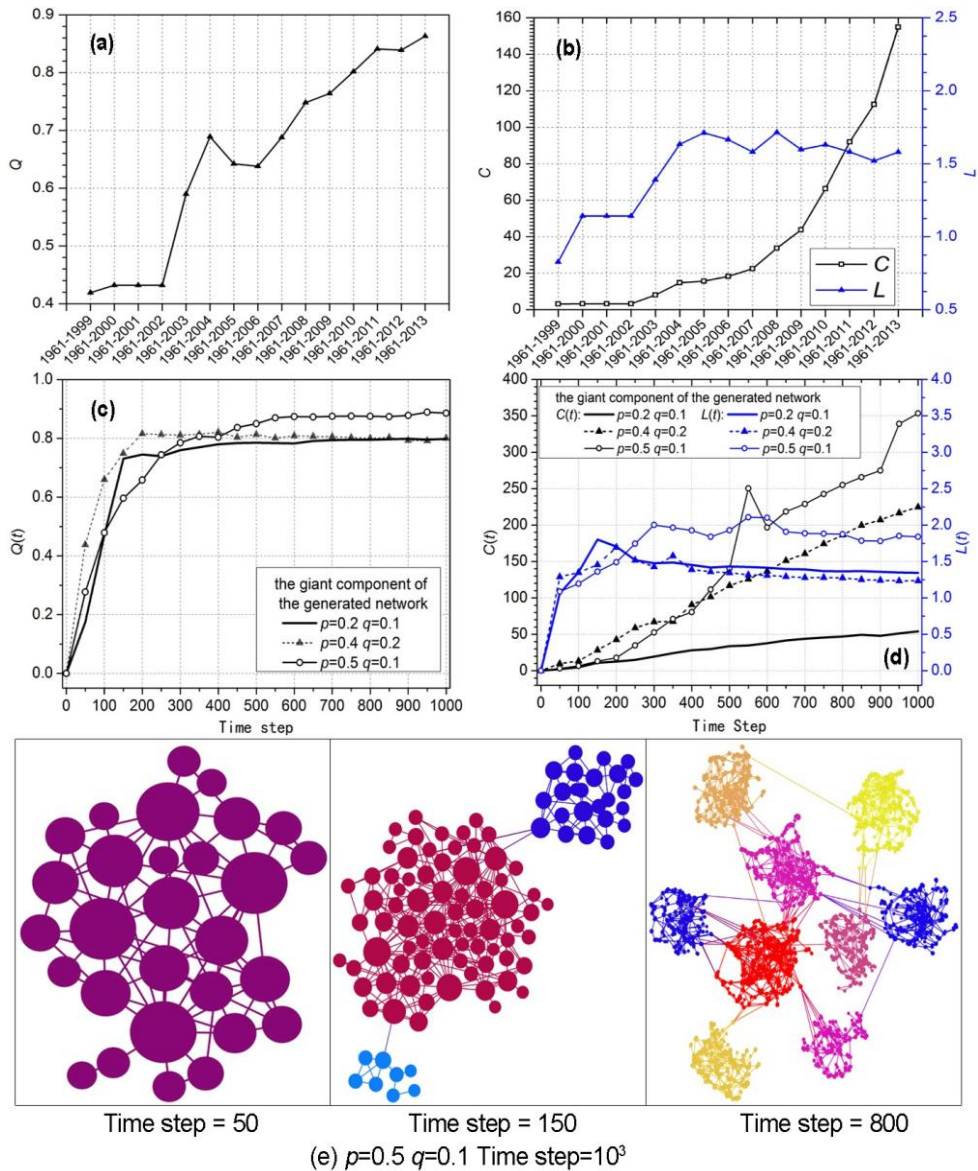

**Figure 5** The results of comparative analysis: (a) modularity of the giant component in the EOC co-authorship network (denoted as $Q$), (b) compared with the random network of same scale, the variations of the clustering coefficient and the average shortest path length of the giant component (respectively denoted as $C$ and $L$) in the EOC co-authorship network, (c) when ($p=0.2$ $q=0.1$), ($p=0.4$ $q=0.2$) and ($p=0.5$ $q=0.1$), the modularity of the giant component in the simulation (i.e. $Q(t)$), (4) compared with the random network of same scale, the variation of clustering coefficient and the average shortest path length of the giant component in the simulation when ($p=0.2$ $q=0.1$), ($p=0.4$ $q=0.2$) and ($p=0.5$ $q=0.1$), (e) in the presented model, the evolution pattern of collaboration network in a run ($p=0.5$ $q=0.1$), the giant component evolves from a small cluster to a chained-community structure, then to a small world network which has clear community structure

*4.2.2 The evolution of the giant component*

Section 4.2.1 reveals the giant component both has significant small-world characteristics and clear community structures in the range of ($0.1 \leq p \leq 0.5$  $0.1 \leq q \leq 0.2$). In this section, three combinations of $p$ and $q$ values in such range (i.e. ($p=0.2$ $q=0.1$), ($p=0.4$ $q=0.2$), ($p=0.5$ $q=0.1$)) are selected as the samples to further explore the evolutionary process of the giant component, and qualitatively compare the evolution of the giant component with the results observed in EOC network. The results of comparative analysis are shown in Figure 5.

Figure 5(a) and (b) depict the changes of modularity (denoted as $Q$), clustering coefficient (denoted as $C$), and average shortest path length (denoted as $L$) of the giant component in the EOC network. The evolution of $Q$, $C$, and $L$ values can be divided into three stages, which reflects the evolutionary process of the giant component in Figure 1. The first stage is from 1999 to 2002. In this stage, the modularity is at a low level ($Q<0.5$). Comparing with the random network of same scale, the clustering coefficient is slightly higher than that of the random network ($C \approx 4.0$), and the average shortest path length is close to that of the random network ($0.75<L<1.25$). The changes of $Q$, $C$ and $L$ in this stage indicate the giant component is just a small cluster which is comprised of closely interlinked nodes. Consequently, the giant component can hardly be divided into sub-communities due to the dense internal links, which also causes high clustering coefficient and small distance between two random nodes in the giant component. The second stage is from 2003 to 2009. In this stage, $Q$ remarkably grows from 0.4 in 2002 to 0.75 reveals the giant component here expands by merging external clusters, resulting in the moderate growth of $C$ ($8<C<44$). Meanwhile, after a sharp increase, $L$ value fluctuates between 1.50 and 1.75 during 2004-2009. This indicates a small number of inter-community links form in the giant component, which also play a crucial role in the emergence of chained-community structure. An obvious change in the third stage (i.e. from 2009 to 2013) is a steep ascent of $C$ value ($C \rightarrow 180$), while $L$ value is roughly stable and $Q$ continuously increases. Comparing with the higher value of $C$, $L$ value is at a relative low level. These changes of $Q$, $C$ and $L$ reflect the further evolution of the giant component. With the expansion of the giant component by continuous merging external clusters, inter-community links constantly form so that the giant component become increasingly modular, and then evolves into a small world network.

As shown in Figure 5(c) and (d), the evolution of the giant component in the three samples is similar to that described in Figure 5(1) and (2). Because the evolutionary trend of the three samples are similar, here we take the network states of ($p=0.5$ $q=0.1$) as an example to specify the evolution of the giant component in our model. It is the initial stage of the giant component in the first 50 time steps. In this stage, we can observe $Q(t)$, $C(t)$ and $L(t)$ are all at a low level (e.g. (50)$\approx$0.3, $C(50) \approx 5.0$, $L(50) \approx 1.25$). This indicates the giant component is a small cluster comprised of closely interlinked agents. When time step increases from 50 to 300, we can observe the rapid ascending of $Q(t)$ and $L(t)$, and a slight growth of $C(t)$. In this sense, the giant component evolves from a small cluster (e.g. the network state of time step 50 in Figure 5(e)) to a chained-community structure (e.g. the network state of time step 150 in Figure 5(e)). Since time step 300, the modularity increases slightly and its values are at a high level around 0.9 after 550 time steps. The average shortest path length is relative stable ($1.75<L(t)<2.25$) while the clustering coefficient has obtained a rapid growth, e.g. the value of $C(300)$ is 50, and then rapidly increases to 140 at time step 500. Both the changes of $C(t)$ and $L(t)$ indicate a small world network emerges in the giant component. From the topology of the network, we can also observe that the giant component is comprised of multi-communities which are interconnected by sparse edges (e.g. the network state when time step reaches 800 in Figure 5(e)). The above analysis shows that a three-stage evolutionary process of the giant component (i.e. from a small cluster to a chained community structure, then to a small world network) can be obtained both in the generated network and the EOC co-authorship network. This further illustrates the presented model may reflect the evolution of actual networks.

We give a partial explanation for the prior phenomenon observed in the simulation. From the aspect of network structure, homophily is essentially a clustering mechanism to form clusters of homogeneous agents. With expansion of clusters by absorbing other agents of same trait, a small number of high-degree agents appear so that they can create link with the agents of different trait by heterophily mechanism. The giant component thus evolves from a small cluster to chained-community structure. During the further evolution of the giant component, the giant component becomes increasingly modular by merging external clusters. Meanwhile, under the mechanism of heterophily, the communities are interconnected with one another due to the increase of high-degree agents. The giant component thus evolves into a small world network with clear community structure.

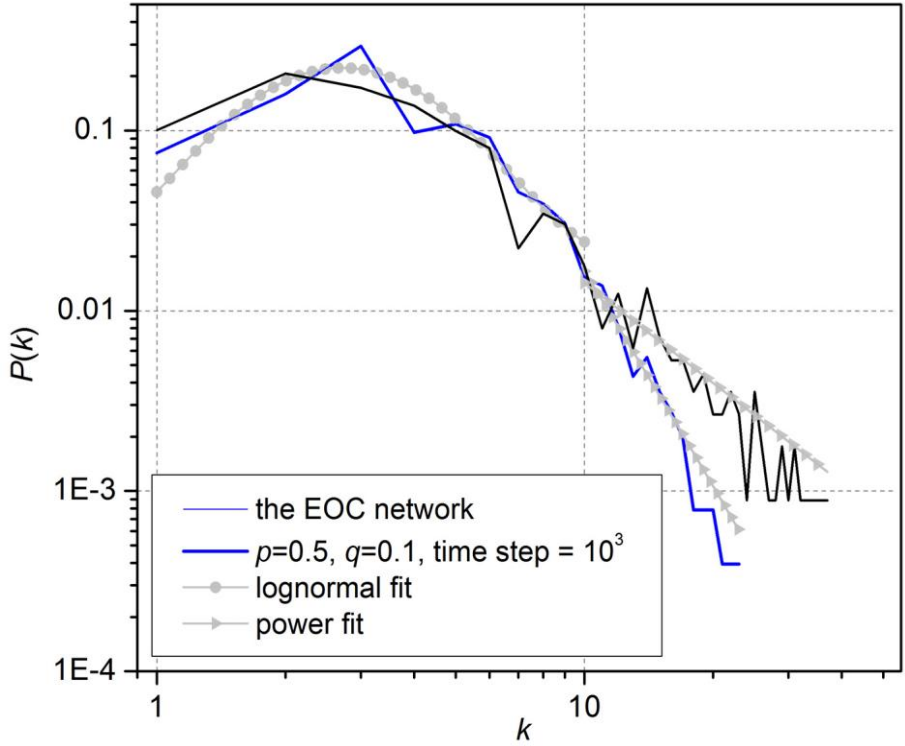

**Figure 6** Fitting of the degree distribution of the giant component in the generated network ($p$=0.5, $q$=0.1 and time step=$10^3$) and that in the EOC co-authorship network (1961~2013), where the power indexes are −1.9 and −4.2 respectively

Besides the similar evolutionary process, the degree distribution in the simulation is also similar to that of the EOC network. Figure 6 respectively plots the degree distribution of the giant component in the simulation (when $p$=0.5, $q$=0.1 and time step reaches $10^3$) and that in the EOC network (1961~2013). In Figure 6, the node degree ($k$) is plotted along the horizontal axis, while the vertical axis displays the probability of the nodes that has a particular degree, denoted by $P(k)$.

The black solid line is the actual degree distribution and the blue solid line is the degree distribution in the simulation, while the gray line with circle is the lognormal fit and the gray line with triangle is power-law fit. Both the degree distributions obey the joint distribution of lognormal and power-law distribution. Most of nodes (i.e. authors or agents) are in the range of $k<10$, in which the degree distribution obeys lognormal distribution. In the range of $k\geq10$, the power-law distribution roughly fits the degree distribution of the giant component, which indicates a small number of nodes have a greater number of collaborators than the average nodes do.

## 5 Conclusion

The community structure and small world characteristic in the scientific collaboration networks inspire extensive discussions on the mechanisms underlying such two structures in academic circles, and related empirical studies show actual collaboration is strongly related to homophily and heterophily in scholars' trait (e.g. research topics and specialties). However, people usually use social capital to explain the formation of community and small world in scientific collaboration networks, i.e. the formation of new collaboration depends on the structure of existing social relationships. The community and small world network thus form through the combination of structural embeddedness and structural holes. In this paper, we argue that the role of preferred-collaboration based on traits should not be overlooked. Especially, some combination of homophily and heterophily in scholars' trait may prompt the scientific collaboration networks to evolve into a small world network which has clear community structures. Therefore, based on our empirical work, we develop a multi-agent model to investigate the evolution of scientific collaboration network under the combinatorial effects of homophily and heterophily. The simulation results can be summarized as follows.

Firstly, with no considerations of social capital, the generated network can also evolve into a modular small world network and the evolutionary process is similar to that observed in the empirical analysis.

Secondly, homophily is essentially a clustering mechanism to form clusters (i.e. the communities in the giant component) of the agents which have similar traits. Heterophily plays an important role in the inter-community links, the establishment of which always requires the involvement of high degree agents.

Thirdly, under a certain combination of homophily and heterophily ($0.1\leq p\leq0.5$ and $0.1\leq q\leq0.2$ in the simulation), the giant component in our model evolves from a small cluster to a chained-community structure, then to a small world network. This evolutionary process is not only consistent with the findings of our empirical work but similar to the evolution of co-authorship network in the field of "complex networks" (Lee et al. 2010). Such result indicates the evolutionary mode obtained by modeling in this paper can reflect the actual evolution of scientific collaboration networks.

The overall results indicate that，besides the structural embeddedness and structural holes, the combination of homophily and heterophily is also a key factor in the emergence of community and small world in the scientific collaboration network. This may be complementary to structural embeddedness and structural holes for understanding the evolutionary dynamics of scientific collaboration network. The evolution of actual networks could be the result of combined effects of structural embeddedness, homophily and heterophily.

To sum up, in this paper we present a primitive attempt to deepen our understandings on the evolutionary patterns of the scientific collaboration networks by a multi-agent model and a few interesting results are obtained. However, the present work is far away from a solid explanation of the social dynamics of scientific collaboration network and the limitations are as follows. Firstly, in the aspect of interactive mechanisms, interaction between individuals will cause the changes of their traits, namely contagion (Christakis and Fowler 2013). In fact, it is worthwhile to examine whether and how such trait-changes caused by contagion affect the evolution of scientific collaboration networks. Secondly, it can be speculated that the evolutionary process in our model may reveal some general characteristics which can also be identified in some other fields due to the consistent results between the simulation and empirical analysis. But only one case is not enough to illustrate these general characteristics in the evolution of scientific collaboration networks. Thus, in the future work, we are going to investigate the scientific collaboration network evolution by further incorporating the mechanism of contagion. Furthermore, we are also to give further examinations of the co-authorship network in other fields, so as to test the generality of the structural evolution observed in our model.